\documentclass[aps,pre,twocolumn,superscriptaddress, showpacs,amsmath]{revtex4}

\usepackage[dvips]{graphicx}

\newcommand{\beq}{\begin{equation}}
\newcommand{\eeq}{\end{equation}}
\newcommand{\beqa}{\begin{eqnarray}}
\newcommand{\eeqa}{\end{eqnarray}}
\newcommand{\kB}{\mbox{$k_{\rm B}$}}
\newcommand{\kBT}{\mbox{$k_{\rm B}T$}}

\begin{document}


\title{
Sticky steps inhibit step motions near equilibrium
}



\author{Noriko Akutsu}
\email[]{nori@phys.osakac.ac.jp}
\affiliation{Faculty of Engineering, Osaka Electro-Communication
 University, Neyagawa, Osaka 572-8530, Japan}


\date{\today}

\begin{abstract}
Using a Monte Carlo method on a lattice model of a vicinal surface with short-range step-step attraction, we show that, at low temperature and near equilibrium, there is an inhibition of the motion of macro-steps.
This inhibition leads to a pinning of steps without defects, adsorbates, or impurities (self-pinning of steps).
We show that this inhibition of the macro-step motion is caused by faceted steps, which are  macro-steps that have a smooth side surface.
The faceted steps result from discontinuities in the anisotropic surface tension (the surface free energy per area).
The discontinuities are brought into the surface tension by the short-range step-step attraction.
The short-range step-step attraction also originates `step-droplets', which are locally merged steps, at higher temperatures.
We derive an analytic equation of the surface stiffness tensor for the vicinal surface around the (001) surface.
Using the surface stiffness tensor, we show that step-droplets roughen the vicinal surface.
Contrary to what we expected, the step-droplets slow down the step velocity due to the diminishment of kinks in the merged steps (smoothing of the merged steps).

\end{abstract}


\maketitle

\section{introduction}

A vicinal surface is a tilted surface deviated from a low Miller-index surface.
The vicinal surface is thought to be described by the terrace-step-kink (TSK) model\cite{beijeren87}-\cite{gmpt} when the temperature is below the roughening transition temperature of the low Miller-index surface.
When the surface steps are regarded as linear excitations, the surface free energy per projected $x$-$y$ area $f(\rho)$\cite{landau,andreev} is obtained as the ground-state energy of the one-dimensional (1D) free fermion (FF) in the following form\cite{beijeren87}-\cite{yamamoto}:
\beq
f(\rho)= f(0)+\gamma \rho + B \rho^3 + {\cal{O} }(\rho^4), \label{frho}
\eeq
where $\rho$ represents the step density, $\gamma$ represents the step tension, and $B$ represents the step-interaction coefficient.
The characteristic feature of Eq. (\ref{frho}) is the lack of a quadratic term for $\rho$.
The form of the free energy of Eq. (\ref{frho}) is common to other many-body systems with nonoverlapping linear excitations embedded in two dimensions (2D)\cite{villain1980,dennij88,lieb}.
The $\rho$-expanded form of the free energy is called the Gruber-Mullins-Pokrovsky-Talapov (GMPT) universal form\cite{gmpt} or the 1D FF universal form\cite{jayaprakash,schultz,akutsu88}.

If we introduce the surface gradient $\vec{p}=(p_x,p_y)=\partial z(x,y)/\partial x, \partial z(x,y)/\partial y)$, $\rho$  is described by $\vec{p}$ as $\rho=|\vec{p}|/d_1$, where $d_1$ (=1) is the step height of an elemental step, and $z(x,y)$ represents the surface height.  
We call $f(\vec{p})$ the vicinal surface free energy.
Using the surface gradient, the surface tension $\gamma_{\rm surf} (\vec{n})$ links to the vicinal surface free energy as 
\beq
\gamma_{\rm surf} (\vec{n})=f(\vec{p})/\sqrt{1+|\vec{p}|^2},
\label{surface_tension}
\eeq
 where $\vec{n}$ is the normal unit vector of the surface at $z(x,y)$.

\begin{figure}
\begin{center}
\includegraphics[width=8 cm,clip]{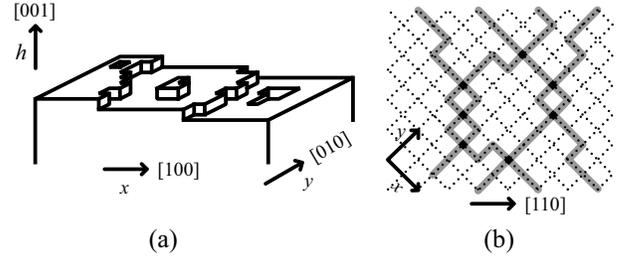}%
\end{center}
\caption{
(a) Perspective view of a surface on the RSOS model tilted towards the $[100 ]$ direction.
(b) Top view of a surface on the RSOS model tilted towards the $[110 ]$ direction.
Gray lines: surface steps.
Filled squares: Collision points of the adjacent steps.
}
\label{vicinal}
\end{figure}

In our previous papers\cite{akutsu10,akutsu11,akutsuJPCM11}, we presented a simple lattice model for a surface with sticky steps, the restricted solid-on-solid  model  with the point-contact-type step-step attraction (p-RSOS model) (Fig. \ref{vicinal}).
Using methods of statistical mechanics, we determined the surface free energy with full anisotropy for the vicinal surface around the (001) surface.
At low temperature, we obtained the anisotropic surface free energy, which included the {\it discontinuities}\cite{akutsu10} due to the step-step attraction. 
In addition, in one temperature region, we obtained non-GMPT values of the shape exponents\cite{akutsuJPCM11} on the profile of the equilibrium crystal shape (ECS), which is the shape of a crystal particle that has the least surface free energy\cite{saito}-\cite{landau},\cite{wulff}-\cite{ookawa}.
The non-GMPT values of the shape exponents obtained by the methods of statistical mechanics closely relate to the non-GMPT $|\vec{p}|$ expanded form of the vicinal surface free energy,  as follows\cite{akutsuJPCM11}:
\beq
f_{\rm eff}(\vec{p})= f(0) + \gamma(\phi) |\vec{p}| + A_{\rm eff}(\phi)|\vec{p}|^2 + B_{\rm eff}(\phi)|\vec{p}|^3+{\cal O}(|\vec{p}|^4). \label{fpprsos}
\eeq
Here we have introduced a polar coordinate system with respect to $\vec{p}$, and we have $p_x=|\vec{p}|\cos \phi$ and $p_y =|\vec{p}|\sin \phi$.
The angle $\phi$ describes the angle of tilt from the $y$-axis, which is formed by the mean running direction of the steps. 
As seen from Eq. (\ref{fpprsos}), a quadratic term with respect to $|\vec{p}|$ has appeared.
In order to obtain Eq. (\ref{fpprsos}), we took the inhomogeneity of the step configuration, the `{\it step-droplets}' (1D boson $n$-mers with finite lifetimes), into consideration\cite{akutsuJPCM11}.
The step-droplets are the locally merged steps in the vicinal surface.
Using Eq. (\ref{fpprsos}), we were able to reproduce the singularity in the surface free energy and the non-GMPT shape exponents, which were obtained with methods of statistical mechanics.

Though there are extensive studies on the mechanisms of step bunching (such as electromigration\cite{stoyanov}-\cite{uwaha}, the Schwoebel effect\cite{pimpinelli,weeks}, the shockwave effect\cite{chernov}, impurity effects\cite{weeks94}-\cite{frank}, and strain effects\cite{teichert}-\cite{ibach}), the vicinal surface free energy is assumed to have the GMPT form (Eq. (\ref{frho})).
A discontinuity in the anisotropic surface tension is expected to strongly affect the morphology of the crystal surface, accompanied by a change in the dynamics of surface steps. 
The combination of the above mentioned mechanisms and a discontinuity in the surface tension will give us a true understanding of the nano- and mesoscale phenomena on the surface.
In addition, since the self-assembling of surface steps is expected to provide new technology to construct nanoscale structures (such as nanowire), the study of the effects of the discontinuity in the surface tension to the morphology and the surface dynamics is important from an industrial point of view.

The aim of this paper, therefore, is to study the effect of the discontinuity in the surface tension to the step dynamics near equilibrium.
We also discuss the morphological aspect of the crystal surface from the viewpoint of the self-assembling of steps.
In order to obtain clear results for the effect of the discontinuity in the surface tension, we exclude realistic interactions on the vicinal surface except for the point-contact-type step-step attraction.

This paper is organized as follows.
In \S II, we present the restricted solid-on-solid (RSOS) lattice model with point-contact-type step-step attraction (the p-RSOS model).
Using a Monte Carlo method with the Metropolis algorithm, we demonstrate the inhibition and the slowing down of the step motion near equilibrium on the p-RSOS model.
In \S III, we show the discontinuous vicinal surface free energy and the equilibrium configuration of the vicinal surface.
We also calculate analytically the surface stiffness tensor near the (001) surface.
In \S IV, the inhibition of the motion of a macro-step, the intermittent motion of the vicinal surface,  and the pinning phenomena in the vicinal surface are studied in connection with `step faceting'\cite{mullins}.
In \S V, we study the roughness of the vicinal surface with step-droplets and the slowing down of the step motion near equilibrium.
Summary and further discussion are given in \S VI, and the conclusion is given in \S VII.

\section{Surface motions near equilibrium}

\subsection{A lattice model for sticky steps}

In this section, we show Monte Carlo results for the surface motion near equilibrium for a vicinal surface with sticky steps. 
The model we adopted for the Monte Carlo calculation is the restricted solid-on-solid (RSOS) model with point-contact-type step-step attraction (p-RSOS model)\cite{akutsu10}-\cite{akutsuJPCM11}.

To describe microscopic surface undulations, let us consider the surface height $h(i,j)$ at a site $(i,j)$ on a square lattice  (Fig. \ref{vicinal}).
In the RSOS model\cite{rsos}, the height differences between nearest-neighbor (nn) sites are restricted to values of $\{1,0,-1\}$.
We consider a point-contact-type microscopic step-step interaction and refer to this model as the p-RSOS model\cite{akutsu10}-\cite{akutsuJPCM11}.
The Hamiltonian for the p-RSOS model can then be written as
\beqa
{\cal H}_{\rm p-RSOS} &=& \sum_{i,j} \epsilon 
[ |h(i+1,j)-h(i,j)|  \nonumber \\
&&
+|h(i,j+1)-h(i,j)|]   \nonumber \\
&& +\sum_{i,j} \epsilon_{\rm int}[ \delta(|h(i+1,j+1)-h(i,j)|,2)  \nonumber \\
&& +\delta(|h(i+1,j-1)-h(i,j)|,2)] ,   \label{hamil}
\eeqa
where $\epsilon$ represents the microscopic ledge energy, $\epsilon_{\rm int}$ is the microscopic step-step interaction energy, and $\delta(a,b)$ is the Kronecker delta.
In the case of $\epsilon_{\rm int}<0$, the interaction between the steps becomes attractive\footnote{The Hamiltonian for this model is similar to the Hamiltonian presented by den Nijs and Rommels\cite{dennij89}, where the step-step interactions are repulsive.}.
The summation with respect to $(i,j)$ is performed over all sites on the square lattice.
The RSOS restriction is required implicitly.

Physically, we consider that the point-contact-type step-step attraction arises from the local formation of a bonding state at the collision point of the adjacent steps.
When adjacent steps collide at a point, the orbital of the dangling bond of each step will overlap, and the spin pairing between electrons in the dangling bonds will form a bonding state\cite{quantumChem}.
The energy gain obtained by forming the bonding state amounts to $-\epsilon_{\rm int}$.

In the p-RSOS model, there are two characteristic temperatures $T_{f,1}$ and $T_{f,2}$ for the vicinal surface tilted towards the $[110 ]$ direction ($\phi= \pi /4$)\cite{akutsuJPCM11}.
At temperatures $T< T_{f,1}$, the surface tension of the (111) surface becomes discontinuous.
At temperatures $T< T_{f,2}$, the surface tension of the (001) surface becomes discontinuous.
That is, on the profile of the ECS, the surface slope of the vicinal surface jumps at the (111) facet edge (a first-order shape transition) for $T< T_{f,1}$, and the surface slope jumps at the (001) facet edge for $T<T_{f,2}$.
Then, the (001) facet directly contacts the (111) facet for $T<T_{f,2}$.

The key concept that is introduced to explain the breakdown of the 1D FF picture is the inhomogeneity resulting from the discontinuity in the surface tension.
In other words, the formation of a `step-droplet' (a 1D boson $n$-mer), which is formed from locally merged steps. 
Using the size of the step-droplet, $A_{\rm eff}(\phi)$ and $B_{\rm eff}(\phi)$ in Eq. (\ref{fpprsos}) are expressed explicitly in the limit of $|\vec{p}| \rightarrow 0$ as follows\cite{akutsuJPCM11}:
\begin{subequations}
\beqa
A_{\rm eff} (\phi) &=&n_0^{(1)} (\phi) \gamma_1^{(1)} (\phi) /d_1, \label{eqAeff} \\
B_{\rm eff} (\phi) &=&  \frac{1}{2d_1} \left[ n_0^{(2)} (\phi) \gamma_1^{(1)} (\phi) + n_0^{(1)} (\phi)^2 \gamma_1^{(2)} (\phi) \right] \nonumber \\
&&
+ \frac{B_1 (\phi) }{d_1^3}   \label{eqBeff},
\eeqa
\end{subequations}
where $d_1$ ($=1$) represents the height of an elementary step, and $B_1$ represents the step-interaction coefficient between the elementary steps.
The variables $n_0^{(m)} (\phi)$, $\gamma_1^{(m)} (\phi)$, and $\tilde{\gamma}_1^{(m)} (\phi)$ are defined by
\begin{subequations}
\beqa
\gamma_1^{(m)} (\phi) &=&\left. \frac{\partial^m (\gamma_n (\phi) /n)}{\partial n^m}\right|_{n=1}, \\
 \tilde{\gamma}_1^{(m)} (\phi) &=& \left. \frac{\partial^m (\tilde{\gamma}_n (\phi) /n)}{\partial n^m}\right|_{n=1}, \\
 n_0^{(m)} (\phi) &=& \left.\frac{\partial^m \langle n (\phi) \rangle }{\partial |\vec{p}|^m}\right|_{|\vec{p}|=0+}  ,
\eeqa
\end{subequations}
where $\gamma_n(\phi)$ represents the step tension of a macro-step, which is formed from the merger of $n$ elementary steps, and $\tilde{\gamma}_n (\phi)  = \gamma_n (\phi)  + \partial^2 \gamma_n (\phi)  / \partial \phi^2$ represents the step stiffness of the merged step. 
$\langle n \rangle$, which describes the mean size of the step-droplets, represents the mean number of elementary steps in a merged step.
Here, $\langle \cdot \rangle $ represents the thermal average.

\subsection{Monte Carlo calculation}

\begin{figure*}
\begin{center}
\includegraphics[width=15 cm,clip]{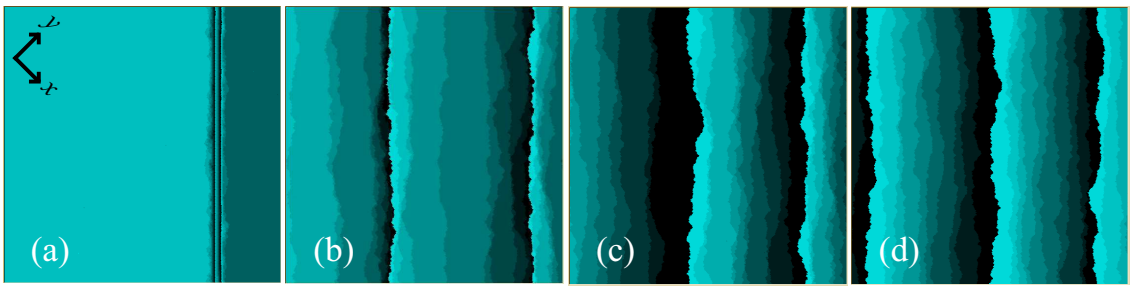}%
\end{center}
\caption{
(Color online) Top views of the vicinal surfaces tilted towards the $[ 110 ]$ direction, calculated by the Monte Carlo method.
The brighter, the higher, with ten gradations. 
$1 \times 10^8$ MCS/site.
$\Delta \mu/\epsilon = 0.0005$.
$\epsilon_{\rm int}/\epsilon = -0.5$
$N_{\rm step}=24$.
Size: $240 \sqrt{2} \times 240 \sqrt{2}$.
(a) A24    ($T< T_{f,2}$). 
(b) B24    ($T_{f,1} <T< T_{f,2}$). 
(c) C24     ($T> T_{f,1}$). 
(d) A24'    (Original RSOS model).
}
\label{035-037}
\end{figure*}

\begin{table}
\caption{\label{tableMClist} Initial values of external parameters for the Monte Carlo calculations. 
Driving force: $ \Delta \mu/\epsilon = \pm 0.0005$.
Size: $240\sqrt{2} \times 240\sqrt{2}$.}
\begin{ruledtabular}
\begin{tabular}{ccccc}
Label& \parbox[t]{1.5cm}{Number of steps} & Temperature &   \parbox[t]{1.5cm}{Surface slope}& \parbox[t]{1.5cm}{Step-step attraction}\\
 & $N_{\rm step}$ & $\kBT/\epsilon$  & $|\vec{p}|$\footnote{$|\vec{p}|= N_{\rm step} /({240\sqrt{2}})$} & $\epsilon_{\rm int}/\epsilon$ \\
\hline
A1 & 1& 0.35 & $2.95 \times 10^{-3}$ & -0.5 \\
A24  & 24& 0.35    & 0.0707  & -0.5\\
B24  & 24  & 0.36    & 0.0707  & -0.5\\
C24  & 24 & 0.37 &  0.0707  & -0.5\\
A240  & 240& 0.35 &   0.707  & -0.5\\
B240  & 240  & 0.36 &   0.707  & -0.5\\
C240  & 240 & 0.37 &   0.707  & -0.5\\
\hline
A1' \footnote{The original RSOS model}  & 1  & 0.35    & $2.95 \times 10^{-3}$  & 0\\
A24'\footnotemark[2]   & 24  & 0.35    & 0.0707  & 0\\
A240'\footnotemark[2]  & 240 & 0.35 &   0.707  & 0\\
\end{tabular}
\end{ruledtabular}
\end{table}

We used the Monte Carlo method to demonstrate the step motion of the nonconserved system\footnote{The nonequilibrium critical dynamics\cite{hohenberg} is known to  follow the behavior of systems with nonconserved order parameter or the behavior of systems with conserved order parameter.}\cite{hohenberg} on a vicinal surface tilted towards the $[110] $ direction (Fig. \ref{035-037}).
Then, the chemical potential difference between the ambient phase and the bulk crystal $\Delta \mu$ becomes an external variable and is the driving force. 

Initially, a macro-step $N_{\rm step}$, made of $N$ steps, was set almost in the middle of a surface with an area of $240 \sqrt{2} \times 240 \sqrt{2} $ (Fig. \ref{035-037}).
The mean surface slope became $|\vec{p}|=N_{\rm step}/(240 \sqrt{2})$.
Periodic boundary conditions were imposed in the vertical direction in Fig. \ref{035-037}.
For the horizontal direction in Fig. \ref{035-037}, the left side of the image was higher than the right side by $N_{\rm step}$.

To study the time evolution of the step configuration, we adopted a simple Metropolis algorithm.
We randomly chose a site $(i,j)$ and allowed its height $h(i,j)$ to increase or decrease with equal probability. Then, if the RSOS restriction was satisfied, the height was updated by the Metropolis algorithm with a probability $P$ described by
\beq
P=\left\{
\begin{array}{ll}
1&   ( \Delta E(i,j)\leq 0 ),  \\
\exp[- \beta \Delta E(i,j)] & ( \Delta E(i,j)>0 ),
\end{array}
\right.
\label{prob}
\eeq
where $\Delta E(i,j)= E(h(i,j)\pm 1)- E(h(i,j)) \mp \Delta \mu$, and $\beta = 1/\kBT$.
The energy $E(h(i,j))$ was calculated using the p-RSOS Hamiltonian shown in Eq. (\ref{hamil}).
The driving force of the motion of the crystal surface was designated by $\Delta \mu$, where $\Delta \mu>0$ for growth and $\Delta \mu<0$ for sublimation.

The values of the parameters for each simulation are shown in Table \ref{tableMClist}.

\subsection{Normal velocity of the surface \label{MCmovements}}

\begin{figure}
\begin{center}
\includegraphics[width=6 cm,clip]{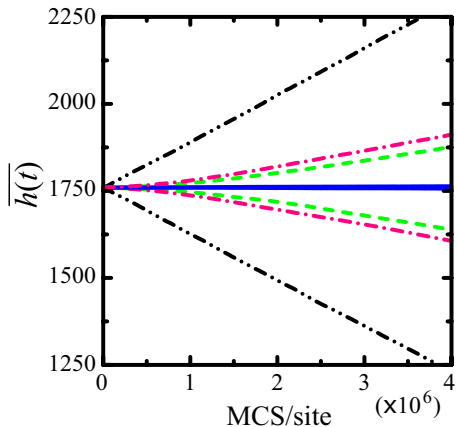}%
\end{center}
\caption{
(Color online) Time evolution of the mean surface height under the p-RSOS model.
Size: $240\sqrt{2} \times 240\sqrt{2}$.
$N_{\rm step}=240$.
$\Delta \mu/\epsilon = \pm 0.0005$.
$\epsilon_{\rm int}/\epsilon = -0.5$.
Full lines: A240, $\kBT/\epsilon=0.35$.
Broken lines: B240, $\kBT/\epsilon=0.36$.
Chain lines: C240, $\kBT/\epsilon=0.37$.
Two-dot chain lines: A240', Original RSOS model ($\epsilon_{\rm int}=0$) at $\kBT/\epsilon=0.35$.
}
\label{time_lag}
\end{figure}

\begin{table}
\caption{\label{tableMC}Monte Carlo results on the surface normal velocities and the lag times. 
Maximum time: $1 \times 10^8 \tau_0$.
$\tau_0$ equals 1 Monte Carlo step/site.
} 
\begin{ruledtabular}
\begin{tabular}{cccc}
Label&  \parbox[t]{2.0 cm}{Normal velocity of the surface\footnote{The velocity with the minus sign corresponds to $  \Delta \mu/\epsilon  = - 0.0005$.}} & Lag time &  \parbox[t]{2cm}{Mean size of step-droplets} \\
 & $ v_z $  ($\times 10^{-6}$ $d_1/\tau_0$) & $\tau_{L}/\tau_0$ ($\times 10^5$)&  $\langle n \rangle$  \\
\hline
A1   & $\pm 0.38 \pm 0.03$ & - & -  \\

A24   & $\pm 0.63 \pm 0.15$ & - & $8.26 \pm 0.02$  \\
B24  & $\pm 7.30 \pm 0.08$& $8 \pm 3 $ & $1.167 \pm 0.003$   \\
C24  & $\pm 7.61 \pm 0.08$& $5 \pm 3$ &$1.102 \pm 0.002$    \\
A240  &  $\pm 0.246 \pm 0.011$ & - & $83.9 \pm 0.6$   \\
B240   &  $\pm 22.90 \pm 0.09$& $22 \pm 4$ & $5.61 \pm 0.08$   \\
C240   &  $\pm 23.87 \pm 0.16$& $7.2 \pm 0.8 $ & $5.28 \pm 0.03$   \\
\hline
A1'   & $\pm 0.381 \pm 0.010$ & - & -  \\
A24' & $\pm 9.07 \pm 0.06$ & $0.4 \pm 0.4$ & $1.00620 \pm 0.00002$    \\
A240'  &  $\pm 66.2 \pm 0.3$ &$0.30 \pm 0.10$ & $6.6220 \pm 0.0003$   \\
\end{tabular}
\end{ruledtabular}
\end{table}

We present snapshots of the vicinal surface near $T_{f,1}$ and $T_{f,2}$ at $\Delta \mu/\epsilon = \pm 0.0005$ in Fig. \ref{035-037}.
From the methods of statistical mechanics\cite{akutsu11,akutsuJPCM11}, we had $\kBT_{f,1}/\epsilon =0.3610 \pm 0.0005$ and $\kBT_{f,2}/\epsilon = 0.3585 \pm 0.0007$, where $\kB$ represents the Boltzmann constant.

In the case of $\kBT/\epsilon=0.35$ ($T < T_{f,2}$) in Fig. \ref{035-037} (a) (A24), the merged step hardly moves forward ($\Delta \mu >0$) or backward ($\Delta \mu <0$).
In the case of $\kBT/\epsilon=0.36$ (B24) and $\kBT/\epsilon=0.37$ (C24), steps bunch locally, and the two cases look similar.
However, the mean terrace width of B24 looks larger than that of C24. As a comparison, we show the case of the original RSOS model ($\epsilon_{\rm int}=0$) in Fig. \ref{035-037} (d) (A24').
From the figure, we see that the mean terrace width of A24' looks smaller than that of C24.

In Fig. \ref{time_lag}, we show the time evolution of the mean surface height $\overline{ h(t)}= (1/{\cal N}) \sum_{i,j} h(i,j) $  for $\Delta \mu /\epsilon = \pm 0.0005$, where $\cal{N}$ represents the number of lattice points.
As seen from Fig. \ref{035-037}, the surface motion is linked to the step movements.

Quantitatively, after the time $1 \times 10^7 \tau_0$, where $\tau_0$ designates the time for one Monte Carlo step per site (MCS/site), the mean surface heights of B24, C24, and A24' increase linearly.
We define the surface velocity $v_z$ as $\partial \overline{h(t)}/\partial t$.
Using the data from the time interval $2 \times 10^7 \tau_0$ to $1 \times 10^8 \tau_0$, we obtained the constant velocities by fitting the data to a linear function with the least-squares method; these velocities are shown in Table \ref{tableMC}.
By extrapolating the linear function to the initial height, we obtained a finite time $\tau_L$, which we call the `lag time'.
The lag times are also shown in Table \ref{tableMC}.

In the case of $\kBT/\epsilon=0.35$ (A24), the center of the macro-step hardly moves.

\section{Discontinuity in the surface tension\label{discontinuity}}

\begin{figure*}
\begin{center}
\includegraphics[width=16 cm,clip]{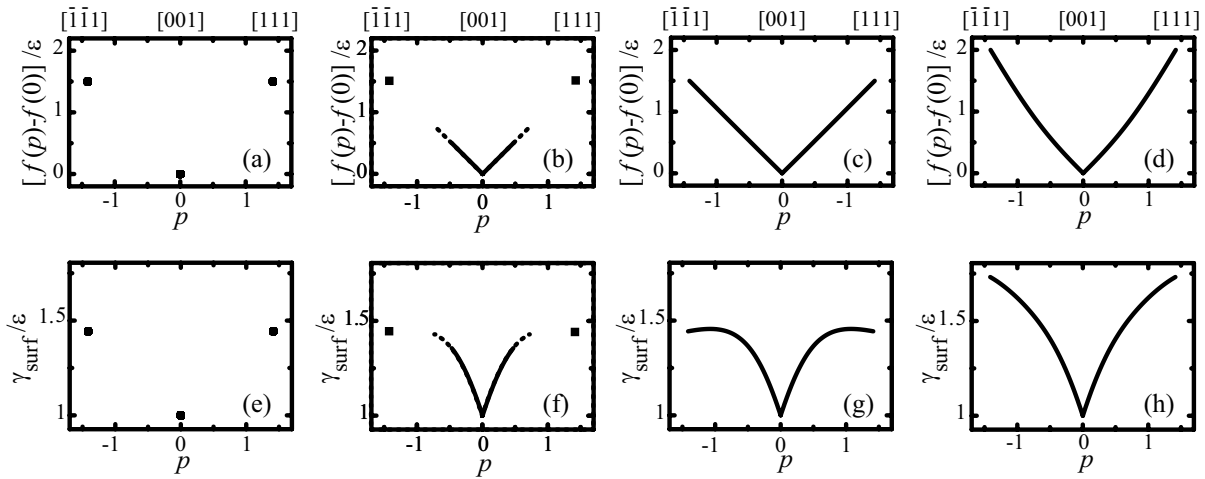}%
\end{center}
\caption{
For the surface tilted towards the $[ 110 ]$ direction calculated by the PWFRG (DMRG) method, (a)-(d) slope dependence of the vicinal surface free energy $[f(p_x,p_y)-f(0,0)]/\epsilon$,   and (e)-(h) surface tension $\gamma_{\rm surf}(p,p)/\epsilon$ .
for (a)-(c) and (e)-(g), $\epsilon_{\rm int}/\epsilon = -0.5$; and .
for (d) and (h), $\epsilon_{\rm int} = 0$, the original RSOS model.
Temperatures:  $\kBT/\epsilon = 0.35$ for (a), (d), (e), and (h); $\kBT/\epsilon = 0.36$ for (b) and (f); $\kBT/\epsilon = 0.37$ for (c) and (g).
Closed squares: (a) and (b), $(0,0)$ and $[f(1,1)-f(0,0)]/\epsilon$; (e) and (f), $\gamma_{\rm surf}(0,0)/\epsilon$ and $\gamma_{\rm surf}(1,1)/\epsilon$. Broken lines: (b) and (f), the curves for the metastable states.
}
\label{fpp035-037}
\end{figure*}

\subsection{Calculations from the methods of statistical mechanics}

Let us consider a vicinal surface that is so close to equilibrium that the diffusivity of atoms does not become the rate-limiting process\cite{pimpinelli93}.
In that case, the interface-limited growth/sublimation becomes the rate-limiting process. 
A detailed understanding of the surface thermodynamic quantities are, therefore, essential for understanding the behavior of a vicinal surface under small $|\Delta \mu|$.
In this section, we study the discontinuity in the vicinal surface free energy $f(\vec{p})$  and the surface stiffness tensor $(f^{ij})$. 

\begin{figure}
\begin{center}
\includegraphics[width=8 cm,clip]{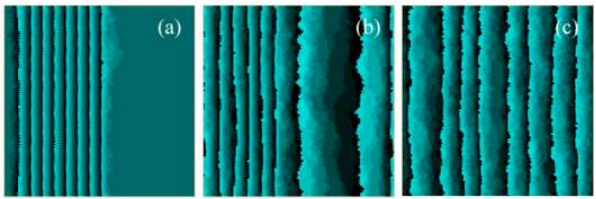}%
\end{center}
\caption{
(Color online) Equilibrium configurations on the p-RSOS model.
Top views of the vicinal surfaces tilted towards the $[110]$ direction calculated by the Monte Carlo method.
The brighter, the higher, with ten gradations. 
$|\vec{p}|=0.707$.
$1 \times 10^8$MCS/site.
$\Delta \mu= 0.0$ (at equilibrium).
$\epsilon_{\rm int}/\epsilon = -0.5$.
$N_{\rm step}=80$.
Size: $80 \sqrt{2} \times 80 \sqrt{2} $.
(a) $\kBT/\epsilon = 0.35$. 
(b) $\kBT/\epsilon = 0.36$. 
(c) $\kBT/\epsilon = 0.37$. 
}
\label{035-037equilibrium}
\end{figure}

In order to calculate the surface free energy with the methods of statistical mechanics, we add the terms of the Andreev field\cite{andreev} $\vec{\eta}= (\eta_x, \eta_y)$ as external variables  to Eq. (\ref{hamil}).
The Andreev field tilts the (001) surface to make a vicinal surface. 
The model Hamiltonian given in Eq. (\ref{hamil}) then becomes
\beqa
{\cal H}_{\rm vicinal}&=&{\cal H}_{\rm p-RSOS} -\eta_x \sum_{i,j}[h(i+1,j)-h(i,j)] \nonumber \\
   && -\eta_y \sum_{i,j}[h(i,j+1)-h(i,j)] .
\label{hamil_vicinal}
\eeqa 
The partition function ${\cal Z}$ for the p-RSOS model is given by
$
{\cal Z}= \sum_{\{h(i,j)\}} \exp[{-\beta {\cal H}_{\rm vicinal}}], 
$
where $\beta= 1/\kBT$. 
The Andreev surface free energy $\tilde{f}(\vec{\eta})$\cite{andreev} is the thermodynamic potential calculated from the partition function ${\cal Z}$ using
\beq
\beta \tilde{f}(\vec{\eta})= - \lim_{{\cal N} \rightarrow \infty} \frac{1}{{\cal N}} \  \ln {\cal Z}, \label{tildef}\\
\eeq
where ${\cal N}$ is the number of lattice points on the square lattice.
The vicinal surface free energy is obtained from the Andreev surface free energy as follows:
\beq
f(\vec{p})= f(\vec{\eta})+ \vec{p} \vec{\eta}.   \label{eq_fpdef}
\eeq

Direct calculation of Eq. (\ref{tildef}) is impractical due to the complexity of the entropy estimations associated with the vast variety of zigzag structures of the surface steps and with the parallel movements of the steps.
Fortunately, the density-matrix renormalization group (DMRG) method\cite{dmrg} developed for the 1D quantum spin system can be used to calculate the partition function Eq. (\ref{tildef}) of the p-RSOS model.
Using the Suzuki-Trotter formula\cite{trotter}, a 1D quantum spin system can be mapped to a transfer matrix\cite{lieb72} of a 2D classical system, such as a surface.
The transfer-matrix version of the DMRG method was developed by Nishino {\it et al.}\cite{pwfrg}-\cite{Ost-Rom} for an infinite lattice, and it is called the product wave-function renormalization group (PWFRG) method.
Since the p-RSOS model can be mapped to a transfer matrix, we adopted the PWFRG method to calculate Eq. (\ref{tildef}).

\subsection{Equilibrium step configurations\label{step-config}}

In Fig. \ref{fpp035-037}, we show the slope dependence of the vicinal surface free energy $f(\vec{p})$ and the surface tension $\gamma_{\rm surf}$ calculated by Eq. (\ref{tildef}),  Eq. (\ref{eq_fpdef}), and Eq. (\ref{surface_tension}), with the PWFRG method.
$f(0,0)$ is assumed to be $\epsilon$, and $\gamma_{\rm surf}(p,p)$ represents $\gamma_{\rm surf}(\vec{n})$ with $\vec{n}=(-p,-p,1)/\sqrt{1+2 p^2}$.

For $\kBT/\epsilon=0.35$ (Fig. \ref{fpp035-037} (a)), only the values of the surface tension of the (001) surface and the (111) surface were obtained.
The surface tension with a mean surface slope in the range $0<|\vec{p}|<\sqrt{2}$ does not exist for the regular train of steps, because the homogeneous surface is thermodynamically unstable there.
Hence, if the mean slope of the vicinal surface has the value of $0<|\vec{p}|<\sqrt{2}$, the surface becomes a mixture of the (001) and (111) surfaces, and its vicinal surface free energy is on the tangent line connecting the value of the (001) surface with the value of the (111) surface. 
We demonstrate this structure in Fig. \ref{035-037equilibrium} (a) using the Monte Carlo method.

For $\kBT/\epsilon=0.36$ (Fig. \ref{fpp035-037} (b)), the surface tension and the vicinal surface free energy increase continuously from $\gamma_{\rm surf}(0,0)=f(0,0)$ as $p$ increases.
Thus, for small $|\vec{p}|$ or small $N_{\rm step}$, a homogeneous structure is expected.
The homogeneous structure (Fig. \ref{035-037} (b)), however, is not the same as the regular train of steps seen in the 1D FF universal system (Fig. \ref{035-037} (d)).
The value of $\langle n \rangle$ is larger than one (see Table \ref{tableMC}), which means that step-droplets are formed.

For large $|\vec{p}|$, $\gamma_{\rm surf}(p,p)$ and $f(p,p)/\epsilon$ jump from $p_t=0.349 \pm 0.002$ to $p_0=1$ at equilibrium.
The broken line in the figure indicates the metastable state for $0.349< p < 0.501$.
The tangent line with the end point $f(1,1)$ contacts the $f(p,p)$ curve at $f(p_t,p_t)$.
Hence, a vicinal surface with mean surface slope $p_t<p<1$ should be formed by the mixture of the surface with the slope $p=p_t$ and the (111) surface.
This structure is demonstrated in Fig. \ref{035-037equilibrium} (b).

For $\kBT/\epsilon=0.37$ (Fig. \ref{fpp035-037} (c)), the surface tension and the vicinal surface free energy are continuous for all $p$.
Therefore, the mixture of two surfaces does not occur in equilibrium (Fig. \ref{035-037equilibrium} (c)).
Step-droplets appear, however, due to the sticky character of the steps.

\subsection{Surface stiffness tensor\label{stiffness}}

Let us describe a slowly undulating surface by $z(x,y)$, where $z(x,y)$ is the surface height at a point $(x,y)$.
A slowly undulating surface is a surface whose orientation varies slowly from place to place around (001).

 Near equilibrium, the time derivative of the height of the slowly undulating surface is assumed to equal the variational derivative multiplied by a transport coefficient ${\cal V}(\{ \vec{p}\} )$\cite{muller-1,kawasaki} based on the time-dependent Ginsburg-Landau theory.
After some manipulation (Appendix \ref{mobility}), we have the following equation:
\beq
v_z= {\cal V}(\{ \vec{p}\} ) \left\{\frac{\Delta \mu}{\Omega}+ \sum_{i,\nu}[f^{i,\nu }z^{(2)}_{\nu,i }] \right\} ,\label{vz2}
\eeq
where $\Omega$ represents the volume of the growth unit in the crystal, $(f^{ij})$ represents the surface stiffness tensor $\partial^2 f(\vec{p})/\partial p_i \partial p_j|_{\vec{p}= \vec{p}_e}$ ($\vec{p}=(p_x,p_y)=(p_1,p_2)$)\cite{akutsu87-1},  and  $z^{(2)}_{ij}$ represents $\partial^2 z/\partial x^i \partial x^j$ ($\vec{x}=(x,y)=(x^1,x^2)$).
Here, $\vec{p}_e$ is the surface slope at equilibrium, and $\partial f(\vec{p})/\partial p_i|_{\vec{p}=\vec{p}_e} =0$ ($i=\{1,2\}$).

The second term on the right-hand side of Eq. (\ref{vz2}) expresses the $x$-$y$ projected Gibbs-Thomson effect.
From the condition $v_z=0$, we obtain another expression of the ECS, except for the facets.

For $T>T_{f,2}$, $\kBT/\epsilon=0.36$, and $\kBT/\epsilon=0.37$,  the surface free energy $f(\vec{p})$ increases monotonically as $|\vec{p}|$ increases around $|\vec{p}|=0$ (Fig. \ref{fpp035-037} (b) and (c)).
We can then calculate the surface stiffness tensor $(f^{ij})$ in Eq. (\ref{vz2}) explicitly in the limit of $|\vec{p}| \rightarrow 0$.

Keeping the non-GMPT $|\vec{p}|$ expanded form of the vicinal surface free energy (Eq. (\ref{fpprsos})) in mind, we describe the  vicinal surface free energy as follows:
\beq
f(\vec{p})= f(0) +\gamma(\phi) \frac{|\vec{p}|}{d} + B_{\zeta}(\phi)\frac{|\vec{p}|^{\zeta}}{d^{\zeta}} + O(|\vec{p}|^{\zeta+1}),
\label{extend_f}
\eeq
where $d$ represents the unit height of a single step ($d=1$), $\gamma(\phi)$ represents the step tension of a single step, and $B_{\zeta}(\phi)$ represents the coefficient of $|\vec{p}|^{\zeta}/d^{\zeta}$.

Adopting Eq. (\ref{extend_f}), and after some calculations, we obtain the expressions of the surface stiffness tensor $(f^{ij})$\cite{akutsu87-1} in the limit of $|\vec{p}| \rightarrow 0$ as follows:
\begin{subequations}
\beqa
f^{11}&=&  \frac{\tilde{\gamma}(\phi)}{|\vec{p}|} \sin^2 \phi + |\vec{p}|^{\zeta-2} (t_1-t_2 \sin ^2 \phi \nonumber \\ &&
 - 2 t_3 \sin \phi \cos \phi) +{\cal O}(|\vec{p}|^{\zeta-1}),
  \label{f11}  \\
f^{22}&=&   \frac{\tilde{\gamma}(\phi)}{|\vec{p}|} \cos^2 \phi   + |\vec{p}|^{\zeta-2} (t_4+t_2 \sin ^2 \phi \nonumber \\ && 
 + 2 t_3 \sin \phi \cos \phi) +{\cal O}(|\vec{p}|^{\zeta-1}),
 \label{f22} \\
f^{12}&=&  - \frac{\tilde{\gamma}(\phi)}{|\vec{p}|} \sin \phi \cos \phi + |\vec{p}|^{\zeta-2} [t_3 (1-2 \sin^2 \phi) \nonumber \\ && 
 + t_2 \sin \phi \cos \phi] +{\cal O}(|\vec{p}|^{\zeta-1})
  \nonumber \\
&=&  f^{21},
 \label{f12} 
\eeqa
\end{subequations}
where $\tilde{\gamma}(\phi)$ ($\tilde{\gamma} (\phi)  = \gamma (\phi)  + \partial^2 \gamma (\phi)  / \partial \phi^2$) represents the step stiffness of a single step, and $t_1$, $t_2$, $t_3$, and $t_4$ are as follows:
\beq
\left.
\begin{array}{lll}
t_1&=&\zeta(\zeta-1) B_\zeta (\phi), \\ 
t_2&=&\zeta (\zeta-2) B_\zeta (\phi)  -B_\zeta'' (\phi), \\
t_3&=& (\zeta-1) B_\zeta' (\phi), \\ 
t_4&=& \zeta B_\zeta (\phi) + B_\zeta'' (\phi) ,\\
B_\zeta'' (\phi)&=&\partial^2 B_\zeta (\phi) /\partial \phi^2 . 
\end{array}
\right\}
\eeq
The principal values of the $(f^{ij})$ become
\begin{subequations}
\beqa
f_n&=&|\vec{p}|^{\zeta-2} t_1 + {\cal O}(|\vec{p}|^{\zeta-1}),  \label{fprincipal}\\
f_t&=& \frac{\tilde{\gamma}(\phi)}{|\vec{p}|} + |\vec{p}|^{\zeta-2} t_4 + {\cal O}(|\vec{p}|^{\zeta-1}). \label{fprincipal_t}
\eeqa
\end{subequations}
Then, we obtain $\det (f^{ij}) $ as follows,
\beqa
\det (f^{ij})&=&  t_1 |\vec{p}|^{\zeta-3}\tilde{\gamma} (\phi)  +{\cal  O}(|\vec{p}|^{\zeta-2}) \nonumber \\ 
&=& |\vec{p}|^{\zeta-3}\zeta (\zeta-1) B_{\zeta}(\phi)\tilde{\gamma}(\phi)  \nonumber \\ 
&& \quad (|\vec{p}| \rightarrow 0). \label{detf} 
\eeqa
Physically, $f_n$ represents the surface stiffness against a bending stress that is normal to the facet edge, and $f_t$ represents the surface stiffness against a bending stress that is tangent to the facet edge.

Using $f_n$ and $f_t$, we have an expression for the normal surface velocity, as follows:
\beq
v_z= {\cal V}(\vec{p})\left\{\frac{\Delta \mu}{\Omega}+ f_n z^{(2)}_n +f_t z^{(2)}_t \right\}, \label{vz3}
\eeq
where 
 $z^{(2)}_n$ and $z^{(2)}_t$ are the eigenvalues of $(\partial^2 z/\partial x^i \partial x^j)$ ($i,j={1,2}$).
According to recent developments in the study of nonequilibrium bunched steps\cite{stoyanov98-1}-\cite{misba10}, the profile of a bunched step is related to the force range of the effective step-step interactions on the nonequilibrium vicinal surface.

At equilibrium, where $v_z=0$, the $z^{(2)}_n$ and $z^{(2)}_t$ are obtained explicitly as 
\beqa
z^{(2)}_{n,eq} &=& -[\Delta \mu/(2 \Omega)] |\vec{p}|^{2-\zeta}/[\zeta(\zeta-1) B_\zeta (\phi)],  \nonumber \\ 
z^{(2)}_{t,eq} &=&  - [\Delta \mu/(2 \Omega)] |\vec{p}|/\tilde{\gamma}(\phi)
\eeqa
 in the limit of $|\vec{p}| \rightarrow 0$.
Therefore, the Gaussian curvature $K$ near the (001) facet edge on the ECS is obtained as follows:
\beq
K = \frac{\lambda^2}{g^2 \det(f^{ij})} 
 \approx \frac{\lambda^2 |\vec{p}|^{3-\zeta}}{ \zeta (\zeta-1) B_{\zeta}(\phi)\tilde{\gamma}(\phi)}, \quad (|\vec{p}| \rightarrow 0) \label{gauss}
\eeq 
where $g=1+|\vec{p}|^2$,  and $\lambda=\Delta \mu/(2 \Omega)$.

Therefore, for $T_{f,2}< T < T_{f,1}$, applying the form of vicinal surface free energy  (Eq. (\ref{fpprsos})) to Eq. (\ref{extend_f}) and  Eq. (\ref{fprincipal})-Eq. (\ref{gauss}),  we have the principal values and the determinant of the stiffness tensor and the Gaussian curvature as follows: 
\begin{subequations}
\beqa
&& f_n=2A_{\rm eff}(\phi), \quad f_t=\tilde{\gamma}_1(\phi)/|\vec{p}|\\
&& \det (f^{ij})=2A_{\rm eff}(\phi)\tilde{\gamma}_1(\phi)/|\vec{p}|\label{detfij036}\\
&& K=\lambda^2|\vec{p}|/[2A_{\rm eff}(\phi)\tilde{\gamma}(\phi)]
\eeqa
\end{subequations}
where $\zeta=2$\cite{akutsuJPCM11}.

For $T_{f,1}<T <<T_R$, where $\zeta=3$\cite{akutsuJPCM11} and $T_R$ is the roughening transition temperature of (001) surface (\S \ref{roughness}),  we have
\begin{subequations}
\beqa
&& f_n=6B_{\rm eff}(\phi)|\vec{p}|, \quad f_t=\tilde{\gamma}_1(\phi)/|\vec{p}|\\
&& \det (f^{ij})=6B_{\rm eff}(\phi)\tilde{\gamma}_1(\phi)/|\vec{p}|\label{detfij037}\\
&& K=\lambda^2/[6B_{\rm eff}(\phi)\tilde{\gamma}(\phi)].
\eeqa
\end{subequations}
Due to the step-droplets, $B_{\rm eff}(\phi)$ (Eq. (\ref{eqBeff})) has a different value from the value of the 1D FF\cite{akutsu88}  in the following:
\beq
B_1(\phi)=(\kBT \pi)^2/[6\tilde{\gamma}(\phi)]. \label{univ_B1}
\eeq
For $T_{f,1}<<T< T_R$, the p-RSOS system converges to the 1D FF system; {\it i.e.}, $B_{\rm eff}(\phi)$ converges to $B_1(\phi)$. 
Hence, $\det (f^{ij})$ and $K$ converge to the GMPT universal value\cite{akutsu88} of 
\beq
\det (f^{ij})= (\kBT\pi)^2 \quad {\rm and} \ K=\lambda^2/(\kBT \pi)^2,\label{detfijFF}
\eeq
respectively.

\begin{table}
\caption{\label{table_pwfrg}Surface stiffness and Gaussian curvature. 
}
\begin{ruledtabular}
\begin{tabular}{ccccc}
 Lable & $\beta f_n$\footnote{$\beta= 1/\kBT$. Eq. (\ref{fprincipal})} & $\beta f_t$\footnote{Eq. (\ref{fprincipal_t})} & $\beta^2 \det(f^{ij})$\footnote{Eq. (\ref{detf})} & $K/(\beta \lambda)^2$\footnote{Eq. (\ref{gauss})} \\
\hline
B24  & $1.24 \times 10^{-2} $ & $19.4$ & $0.241$ & $4.10$  \\
C24  & $0.119$& $19.3$ & $2.30$ & $0.431$   \\
\hline
A24'  & $0.507$ & $19.5$ & $\pi^2\footnote{GMPT (1D FF) universal value, ref.\cite{akutsu88}} $ & $1/(g \pi)^2$\footnote{$g=1+|\vec{p}|^2$. $|\vec{p}|=0.0707.$} \\
  &  &  & $ \approx 9.870$ & $\approx 0.1003$  \\
\end{tabular}
\end{ruledtabular}
\end{table}

In Table \ref{table_pwfrg}, we show the values of the surface thermodynamic quantities for $\kBT/\epsilon=0.36$ (B24) and $\kBT/\epsilon=0.37$ (C24).
As a comparison, we show the values of the original RSOS model at $\kBT/\epsilon=0.35$ (A24').
In the process of the estimation of $f_n$ and $f_t$, we used values calculated by the PWFRG method for $\beta A_{\rm eff}(\pi/4)$ and $\beta B_{\rm eff}(\pi/4)$\cite{akutsuJPCM11}; {\it i.e.}, $\beta A_{\rm eff}(\pi/4)= (6.22 \pm 0.06) \times 10^{-3}$ for B24 and $\beta B_{\rm eff}(\pi/4) = 0.281 \pm 0.008$ for C24.
We also approximated $\beta \tilde{\gamma}_1(\pi/4)$ by the values of the interface stiffness of the 2D nn Ising model\cite{akutsu86,rottman81,akutsu90}; {\it i.e.}, 1.377, 1.371, and 1.364 for A24', B24, and C24, respectively.

As seen from Table \ref{table_pwfrg}, even though $f_t$ is kept almost constant, $f_n$ changes drastically as the temperature changes.
This means that the step-droplets soften the surface against the bending force normal to the mean running direction of the steps.
In fact, the Gaussian curvature at the (001) facet edge increases.

Note that $\vec{p}$ on the ECS ($z=z(x,y)$) depends on $(x,y)$.
Taking the $x$-axis normal to the edge of the (001) facet at a point $(x_c,0)$, \beq
\beta \lambda [z(x,0)-z(x_c,0)]= -{\cal A}_n [\beta \lambda (x-x_c)]^{\theta_n},
\eeq
where $\beta = 1/\kBT$, $\theta_n$ is the normal shape exponent, and ${\cal A}_n$ is the normal amplitude.
Using Eq. (\ref{extend_f}), $\theta_n$ and ${\cal A}_n$ are then described as follows\cite{akutsuJPCM11}:
\begin{subequations}
\beqa
\theta_n &=& \frac{\zeta}{\zeta-1}, \quad {\cal A}_n=\frac{1}{\theta_n}\left[ \frac{\kBT}{\zeta B_{\zeta}} \right] ^{\theta_n-1}, \\
|\vec{p}|&=&\left | \frac{\lambda (x-x_c)}{\zeta B_{\zeta}}\right | ^{\theta_n-1}. 
\eeqa
\end{subequations}

\section{\label{faceting}Inhibition of step motion due to `step faceting'}

\subsection{Roughness on the side surface of a merged step}

In the section, we explain the inhibition of the macro-step motion in the vicinal surface at sufficiently low temperatures $T<T_{f,2}$.

For $T<T_{f,2}$, the inhomogeneous vicinal surface, a mixture of the (001) and (111) surfaces (Fig. \ref{035-037equilibrium} (a)), is realized due to discontinuities in the surface tension\cite{akutsuJPCM11}.
Since both the (001) surface and the (111) surface are smooth, a macro-step in the vicinal surface becomes a faceted step (`step faceting'\cite{mullins}).
Hence, the squared surface widths for the (001) and (111) surfaces should both be finite.

\begin{figure}[htbp]
\begin{center}
\includegraphics[width=5.5 cm,clip]{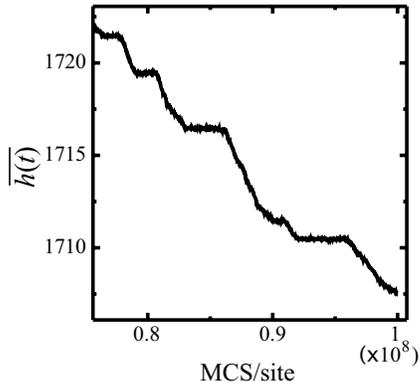}%
\end{center}
\caption{
Intermittent sublimation of the surface. 
A24.
$\Delta \mu/\epsilon = - 0.0005$.
$\epsilon_{\rm int}/\epsilon = -0.5$.
$\kBT/\epsilon = 0.35$. 
$N_{\rm step}=240$.
Size: $240 \sqrt{2} \times 240 \sqrt{2} $.
}
\label{intermittent}
\end{figure}

The squared surface width of a vicinal surface $W(\vec{n})^2$ is defined as follows\cite{akutsu87-1}:
\beq
W(\vec{n})^2= \langle [z(\vec{x})- \langle z(\vec{x}) \rangle ]^2 \rangle /g, \label{Wdef}
\eeq
where $\langle \cdot \rangle $ represents the thermal average and $\vec{n}=(-p_1,-p_2,1)/g$ represents the normal unit vector of the tilted surface.
From the result of Ref.\cite{akutsu87-1}, $W(\vec{n})^2$ is connected to the surface stiffness tensor in the limit of $L \rightarrow \infty$, where $L$ is the linear size of the surface, as follows:
\beq
W(\vec{n})^2=\frac{\kBT}{2 \pi g \sqrt{\det(f^{ij})}}\cdot \ln L . \label{lnL}
\eeq
Since $W(\vec{n})^2$ should be finite for a smooth surface, $\det(f^{ij})$ should be  divergent in the order of $(\ln L)^2$  in the thermodynamic limit.
Hence, slight deformations from the flat surfaces of (001) or (111) are pulled back to the flat surfaces because of the strong Gibbs-Thomson effect.

Further, if we consider $z^{(2)}_n=0$ and $z^{(2)}_t=0$ on the macroscopic vicinal surface, the transport coefficient ${\cal V}(\vec{p})$ reduces to zero because the kink density on the side of the macro-step converges to zero in the thermodynamic limit.
Therefore, the continuous motion of the surface described by Eq. (\ref{vz2}) or Eq. (\ref{vz3}) is inhibited in this temperature region.

Instead, as shown in Fig. \ref{intermittent}, the intermittent motion of the surface occurs in a longer time scale.
The mechanisms that cause the intermittent surface motion are considered to be 2D nucleation\cite{saito} around the intersection line of two surfaces, a terrace surface and the side surface of a faceted step.
Near equilibrium, the nucleation rates at about the center of the terrace and at about the center of the side surface are small, and the nucleation rate around the intersection line of the surfaces becomes larger because of the special geometrical arrangement.

\subsection{\label{self-pinning}Self-pinning of steps}


\begin{figure}[htbp]
\begin{center}
\includegraphics[width=8 cm,clip]{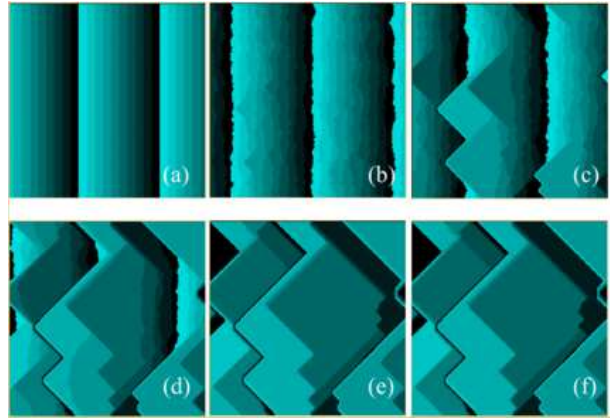}%
\end{center}
\caption{
(Color online) Self-pinning of steps on the p-RSOS model.
Top views of the vicinal surfaces tilted towards the $[110]$ direction calculated by the Monte Carlo method.
The brighter, the higher, with ten gradations. 
$\Delta \mu/\epsilon = 0.1$.
$\epsilon_{\rm int}/\epsilon = -0.5$.
$\kBT/\epsilon = 0.1$. 
$N_{\rm step}=24$.
Size: $240 \sqrt{2} \times 240 \sqrt{2} $.
(a) Initial configuration.
(b) $1000$ MCS/site.
(c) $2000$ MCS/site.
(d) $3000$ MCS/site.
(e) $5000$ MCS/site.
(f) $1 \times 10^4$ MCS/site.
}
\label{pinning}
\end{figure}

\begin{figure}[htbp]
\begin{center}
\includegraphics[width=8 cm,clip]{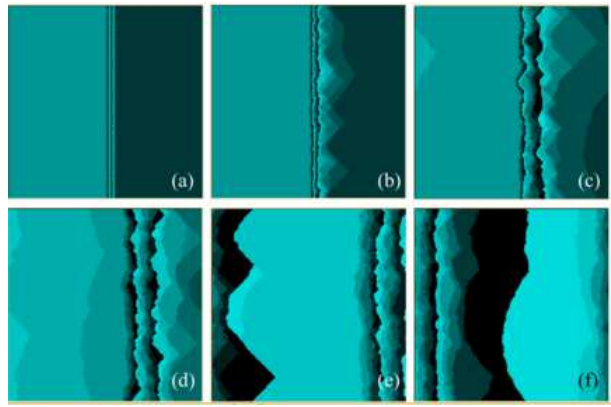}%
\end{center}
\caption{
(Color online) Step growth on the p-RSOS model.
Top views of the vicinal surfaces tilted towards the $[110 ]$ direction demonstrated by the Monte Carlo method.
The brighter, the higher, with ten gradations. 
$\Delta \mu/\epsilon = 0.35$.
$\epsilon_{\rm int}/\epsilon = -0.5$
$\kBT/\epsilon = 0.1$. 
$N_{\rm step}=24$.
Size: $240 \sqrt{2} \times 240 \sqrt{2} $.
(a) Initial configuration.
(b) $1000$ MCS/site.
(c) $3000$ MCS/site.
(d) $5000$ MCS/site.
(e) $9000$ MCS/site.
(f) $1.3 \times 10^4$ MCS/site.
}
\label{growth}
\end{figure}

At sufficiently low temperature, the step movements of the macro-steps are almost inhibited for small $\Delta \mu$ because of step faceting.
In Fig. \ref{pinning}, we show the time evolution of the vicinal surface at $\kBT/\epsilon = 0.1$ for the p-RSOS model with $\epsilon_{\rm int}/\epsilon=-0.5$, starting from the regular train of elementary steps (Fig. \ref{pinning} (a)).
In our previous paper\cite{akutsu11}, we showed that for $\Delta \mu=0$, step zipping, the phenomena of successive sticking of adjacent steps, much like a slide fastener in clothes, occurs at the collision points of adjacent steps.

In the case of $\Delta \mu/\epsilon =0.1$, step-flow growth occurs in the early stages (Fig. \ref{pinning} (b)) of surface growth.
When an elementary step collides with an adjacent elementary step, the steps merge to form a double step due to the sticky character of steps (Fig. \ref{pinning} (b)).
Since the step-droplet now has a velocity lower than that of the elementary steps, the steps from behind catch up with the new step-droplet and merge to form a larger step-droplet (Fig. \ref{pinning} (c)). 
Since the larger step-droplet now has even lower velocity, step movements become pinned by step-droplets, as in a traffic jam (Fig. \ref{pinning} (c)-(e)). 
The mobility of the merged steps is so slow that the vicinal surface is almost quenched when elementary steps disappear (Fig. \ref{pinning}(e),(f)).
This inhibition of the step motion occurs when $\Delta \mu/\epsilon < 0.3$ and around $\Delta \mu/\epsilon = 0.3$ if $\kBT/\epsilon=0.1$.

At $\Delta \mu/\epsilon =0.35$, elementary steps begin to separate successively from the lower side of the macro-steps (Fig. \ref{growth}).
This separation of elementary steps looks like step nucleation.
As mentioned at the end of the previous subsection, step nucleation starts from the 2D nucleation around the intersection line between a terrace surface and a side surface of a macro-step.
In addition, the growing steps in Fig. \ref{growth} exhibit distinctive wavy shapes due to the sticky character of the steps.
The frequency of the separation of steps is not simply defined, and it may be considered to be related to `kinetic roughening'\cite{nozieres,krug1991,vicsek} on the surface.

This separation of steps occurs more frequently when $\Delta \mu/\epsilon > 0.35$, so we designate $\Delta \mu^*/\epsilon = 0.35$ the threshold driving force for the separation of a step at $\kBT/\epsilon=0.1$. 
When the temperature increases within the range of $T<T_{f,2}$, the value of the threshold force for the separation of step $\Delta \mu^*$ becomes smaller.
This is because a decrease of the step tension for an elementary step, which is caused by the entropy of the zigzag structure in an elementary step, induces more frequent 2D nucleation at the intersection line.

In the case of sublimation ($\Delta \mu<0$), an elementary step appears from the upper side of the macro-step, and the step goes backward.
This separation of elementary steps also looks like `step nucleation', and the steps that move backward exhibit distinctive wavy shapes, which are almost mirror-symmetric to the shapes observed for the growing steps.

\section{\label{slowing-down}Slowing down of the step motion due to step-droplets}

\begin{table}
\caption{\label{table_kink}Surface roughness and surface velocities.   $  \Delta \mu/\epsilon  = \pm 0.0005$. 
}
\begin{ruledtabular}
\begin{tabular}{cccccc}
Label &  $ W(n)^2/\ln L$\footnote{Eq. (\ref{lnL})}  & $\rho_{k,1}$\footnote{Eq. (\ref{rhok1})} & $ \rho_{k,2} $ \footnote{Eq. (\ref{rhok2})}& $\rho_{k}$ \footnote{Eq. (\ref{rhok})} & $v_z$\footnote{Eq. (\ref{nu}) and Eq. (\ref{vz4}) }($\times 10^{-6}$ $d_1/\tau_0$) \\
\hline
A1 & 0 & $0.116$ & - & $0.116$ & $0.346$   \\
B24  &  $0.322$ & $0.118$& $0.0598$ & $0.108$ & $7.49$  \\
C24  & $0.104$ & $0.119$& $0.0622$ & $0.113$ & $7.63$  \\
\hline
A1' & $1/(2\pi^2g)$\footnote{$g=1+|\vec{p}|^2$} &  $0.131$& - & $0.131$ & $0.388 $ \\
 &$\approx 0.05066$ &  &  &  &  \\
A24'  & $1/(2\pi^2g)$ & $0.131$& $0.226$ & $0.131$ & $9.36$  \\
  & $\approx 0.05041$ & &  &  &  \\
\end{tabular}
\end{ruledtabular}
\end{table}


\subsection{\label{roughness}Roughness of the vicinal surface}

In this section, we study the step motion in the case of $T_{f,2}<T \sim T_{f,1}$.
In this temperature range, the surface tension and the vicinal surface free energy $f(\vec{p})$ are discontinuous at the (111) surface but continuous around $|\vec{p}|=0$. 
Hence, Eq. (\ref{vz2}) is applicable, and it describes the motion of the vicinal surface around $|\vec{p}|=0$.
Also, for the mean flat surface, the velocity of the surface $v_z$ is expressed using Eq. (\ref{vz2}) as follows:
\beq
v_z={\cal V}(\vec{p})\Delta \mu/ \Omega \label{vz4}.
\eeq 
According to simple linear response theory\cite{kuboII}, the transport coefficient ${\cal V}(\vec{p})$ should be proportional to the squared fluctuation width $W^2(\vec{n})$ of the vicinal surface due to the fluctuation-dissipation theorem (Appendix \ref{fdtheorem} Eq. (\ref{vz6})).
In this subsection, therefore, we study the roughness of the vicinal surface.

Let us first evaluate the roughening transition temperature $T_R$ of the (001) surface. 
By using the PWFRG method, we calculated $T_R$ from the universal relationship $K_R/\lambda^2=4/(\kBT_R \pi)^2$, where $K_R$ is the GMPT Gaussian curvature on the ECS at $T=T_R$\cite{beijeren87,jayaprakash}.
The obtained $T_R$ are   $\kBT_R/\epsilon = 1.505 \pm 0.008$ for $\epsilon_{\rm int}/\epsilon=-0.5$ and $\kBT_R^{\rm RSOS}/\epsilon = 1.584 \pm 0.006$ for $\epsilon_{\rm int}=0$.
The value of $T_R^{\rm RSOS}$ is consistent with the value obtained by den Nijs\cite{dennijs}.
The result that $T_R < T_R^{\rm RSOS}$ means that the step-step attraction slightly roughens the (001) surface.
The local bonds $\epsilon_{\rm int}$ are considered to stabilize the `blobs'\cite{abraham86} in the zigzag structure of a single step.
These blobs enhance the deformations of a step by thermal fluctuation, which decreases the step free energy per length.

The roughness of the vicinal surface is measured by the squared surface width $W(\vec{n})^2$. 
Hence, by substituting Eq. (\ref{detf}) into Eq. (\ref{lnL}), we have
\beqa
W(\vec{n})^2= \frac{\kBT |\vec{p}|^{\frac{3-\zeta}{2}}}{2 \pi g \sqrt{ \zeta (\zeta-1) B_{\zeta}(\phi)\tilde{\gamma}(\phi)}}\cdot \ln L.
\label{W2b}
\eeqa
Using the values of $\det (f^{ij})$ in Table \ref{table_pwfrg}, we calculated the corresponding values of $W(n)^2/\ln L$ and present them in Table \ref{table_kink}.

For $T_{f,2}<T<T_{f,1}$, we can see from Table \ref{table_kink} that the step-droplets roughen the vicinal surface.
For $T_{f,1}<T<<T_R$, the vicinal surface is still roughened by the remaining step-droplets. 
As the temperature increases from $T_{f,1}$, the roughness of the vicinal surface $W(\vec{n})^2/\ln L$ ($|\vec{p}| \rightarrow 0$) rapidly converges to the GMPT value $1/(2 \pi^2)$.

In summary, for the cases demonstrated in \S \ref{MCmovements}, we have the following relationship from the values in Table \ref{table_kink}: 
\beqa
&&W^2(\vec{n})|_{\hat{T}=0.36} > W^2(\vec{n})|_{\hat{T}=0.37} \nonumber \\
&& > W^2(\vec{n})^{\rm RSOS}|_{\hat{T}=0.37} = W^2(\vec{n})^{\rm RSOS}|_{\hat{T}=0.35}, \label{w2order}
\eeqa
where $\hat{T}$ represents the reduced temperature $\kBT/\epsilon$.

\subsection{Step smoothing\label{smoothing}}

From Eq. (\ref{w2order}), referring to the expression of ${\cal V}(\vec{p})$ (Eq. (\ref{vz6})), we must have the relationship $v_z|_{\hat{T}=0.36} > v_z|_{\hat{T}=0.37} > v_z^{\rm RSOS}|_{\hat{T}=0.35}$ for the cases demonstrated in \S \ref{MCmovements}.
However, the calculated order of surface mobility is clearly seen from Fig. \ref{time_lag} and from Table \ref{tableMC} as 
\beq
v_z|_{\hat{T}=0.36} < v_z|_{\hat{T}=0.37} 
< v_z^{\rm RSOS}|_{\hat{T}=0.35}. \label{vzorder}
\eeq
Therefore, even near equilibrium, the simple linear response theory is not applicable to describe the transport coefficient for the vicinal surface with sticky steps around $T \sim T_{f,1}$.

It is the formation of the step-droplets that prevents the application of the simple linear response theory to the transport coefficient.
Though the step-droplets roughen the vicinal surface, they also diminish the number of kinks on the side of the merged steps.
At temperatures sufficiently lower than $T_R$, the step motion is governed by the total in/out flow of materials at the kink sites.
We will use `step smoothing' to refer to this reduction in the number of kinks.

\begin{figure}[htbp]
\begin{center}
\includegraphics[width=8.0 cm,clip]{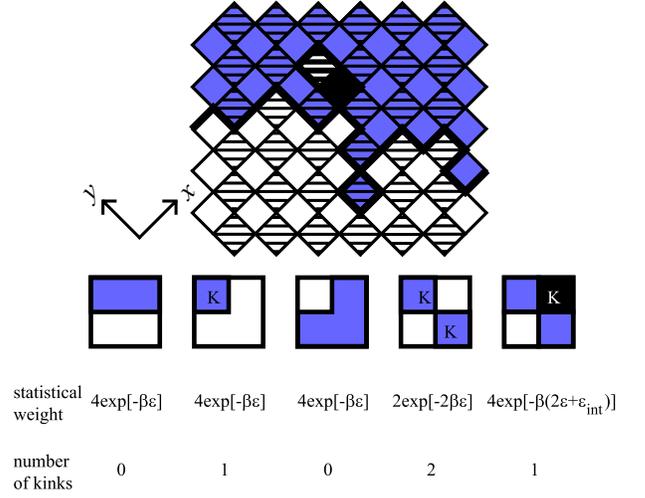}%
\end{center}
\caption{
(Color online) Conceptual diagrams of kink configurations for an elementary step.
 $\Delta \mu>0$.
Top figure: an example of an elementary step (top view).
The lattice sites are divided into two sub-lattices, distinguished by stripes and solid.
Darkly shaded areas describe the lower terrace.
`K' represents a kink site.
}
\label{kink1config}
\end{figure}

We note that the transport coefficient is described as follows:
\beq
{\cal V}(\vec{p}) = \rho_{k} N_{\rm [\bar{1}10]} N_{\rm step}\Omega^2 /(S \kBT), \label{nu}
\eeq
where $\rho_{k}$ represents a kink density, $N_{\rm [\bar{1}10]}$ represents the linear number of the unit cell in the direction of ${\rm [\bar{1}10]}$, and $S=240 \times 240 \times 2$ represents the size of the simulated area.
We show a `kink' in the case of the growth ($\Delta \mu>0$) of a single elementary step in Fig. \ref{kink1config}.
The kink density for a single elementary step is approximated based on the 19-vertex model\cite{rsos} as follows (Fig. \ref{kink1config}):
\beqa
\rho_{k,1} \approx 4\frac{(1+2{\rm e} ^{-\beta \epsilon})^2} {[6 + {\rm e} ^{-\beta \epsilon} + 2{\rm e} ^{-\beta (\epsilon + \epsilon_{\rm int})}]^2}. \label{rhok1}
\eeqa

Using the expressions of $\rho_{k,1}$ and ${\cal V}(\vec{p})$,  we obtained $v_z$ for an elementary step, as shown in Table \ref{table_kink} for the data of A1 and A1'. 
By comparing these values with the ones in Table \ref{tableMC}, it is seen that  the values obtained by Eq. (\ref{nu}) and Eq. (\ref{rhok1}) are in agreement with the values obtained by the Monte Carlo method without fitting parameters.

When $|\vec{p}|$ increases, that is, $N_{\rm step}$ increases, the step-droplets are created by a local merging among steps.
Since $\langle n \rangle$, which represents the mean size of the step-droplets (boson $n$-mers), is close to 1\cite{akutsuJPCM11} for the surface with $N_{\rm step}=24$ (Table \ref{tableMC}), the double step is statistically dominant among multiply merged steps. 
Hence, we consider the mean kink density as follows:
\beqa
\rho_{k} = (2- \langle n \rangle) \rho_{k,1} + (\langle n \rangle-1) \rho_{k,2}, \label{rhok}
\eeqa 
where $\rho_{k,2}$ represents the kink density of the double step.
In equation (\ref{rhok}), we assumed $\langle n \rangle \approx 1\cdot(1-\hat{p}) + 2\cdot \hat{p}$, where $\hat{p}$ represents the fraction of steps that are double steps.

\begin{figure}[htbp]
\begin{center}
\includegraphics[width=8 cm,clip]{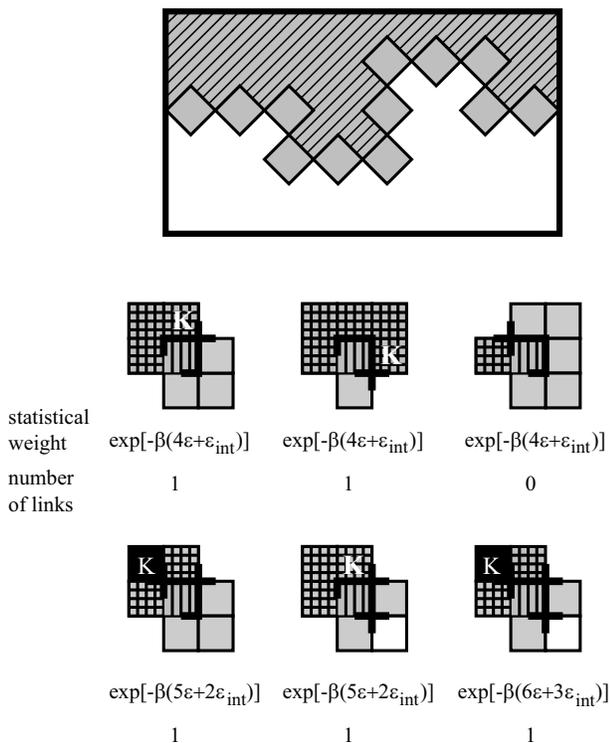}%
\end{center}
\caption{
Conceptual diagrams of kink configurations for a double step.
 $\Delta \mu>0$.
Top figure: an example of a double step (top view).
More darkly shaded areas describe lower terraces: white $>$ gray $>$ gray with stripes $>$ gray with hatching $>$ black.
`K' represents a kink site.
}
\label{kink2config}
\end{figure}

Since the configuration of a double step (Fig. \ref{kink2config}) is approximately described by the triple vertices, $\rho_{k,2}$ is estimated as follows:
\beqa
\rho_{k,2} \approx \frac{[2+ 2 {\rm e} ^{-\beta (\epsilon + \epsilon_{\rm int})}+ {\rm e} ^{-\beta (2\epsilon + 2\epsilon_{\rm int})} ]{\rm e} ^{\beta  \epsilon_{\rm int}}}{3[3 + 2 {\rm e} ^{-\beta (\epsilon + \epsilon_{\rm int})} + {\rm e} ^{-\beta (2\epsilon + 2\epsilon_{\rm int})}]} . \label{rhok2}
\eeqa
In Table \ref{table_kink}, we show the values of $\rho_{k,1}$, $\rho_{k,2}$, and $\rho_k$ calculated by using Eq. (\ref{rhok1}), Eq. (\ref{rhok2}), and Eq. (\ref{rhok}), respectively.
Here, we used the value of $\langle n \rangle$ in Table \ref{tableMC}.
Comparing the values of $v_z$ in Table \ref{table_kink} obtained by Eqs. (\ref{nu})-(\ref{rhok2}) with the ones obtained by the Monte Carlo calculations (\S \ref{MCmovements}), we see that they are in close agreement without fitting parameters.

\section{Summary and discussion}

We studied the step motions in a vicinal surface near equilibrium using a lattice model with sticky steps.
We adopted a restricted solid-on-solid model with a point-contact-type step-step attraction (Eq. (\ref{hamil}), $\epsilon_{\rm int}<0$, the p-RSOS model). 
We showed that the point-contact-type step-step attraction caused discontinuities in the surface tension $\gamma_{\rm surf}(\vec{n})$ (Eq. (\ref{surface_tension})) and in the vicinal surface free energy $f(\vec{p})$ at low temperatures (Fig. \ref{fpp035-037} (a)).
Due to the discontinuities, `step faceting'\cite{mullins} occurred on the macro-steps in the vicinal surface tilted from the (001) surface towards the $[ 110] $ direction for $T< T_{f,2}$ (Fig. \ref{035-037equilibrium} (a)).
The continuous motion of the macro-steps was inhibited under a small driving force $\Delta \mu$ by the step faceting (Fig. \ref{035-037} (a), Fig. \ref{time_lag}).
Instead, the intermittent motion of the vicinal surface (Fig. \ref{intermittent}, Fig. \ref{growth}) occurred by way of 2D nucleation around the intersection line between the terrace surface and the side surface of the macro-step.
Starting from the regular train of steps as the initial configuration, pinning of steps was demonstrated to take place from the collision point of the adjacent elementary steps (Fig. \ref{pinning}) without impurities or defects.
We term this phenomenon the `self-pinning' of steps.

For $T_{f,2}<T < T_{f,1}$, the surface tension and the vicinal surface free energy were continuous for the surface slope $|\vec{p}| \sim 0$, but they were still discontinuous around the (111) surface (Fig. \ref{fpp035-037} (b)).
This discontinuity around the (111) surface led to the formation of `step-droplets' (boson $n$-mers), locally merged steps.
The step-droplets roughen the vicinal surface, and we showed this by calculating the squared surface width $W^2(\vec{n})$ (\S \ref{roughness}, Table \ref{table_kink}), using the relationship between the squared surface width and the determinant of the surface stiffness tensor $\det (f^{ij})$ (\S \ref{stiffness}, Table \ref{table_pwfrg}).
The step velocity, which is expected to be larger than that of the original RSOS model,  was lower than the step velocity of the original RSOS model at the same temperature (Fig. \ref{time_lag}, Table \ref{tableMC}).
This is because the step-droplets diminish the kink density of the vicinal surface (the `step smoothing', \S \ref{smoothing}).
We estimated the transport coefficient (Eq. (\ref{nu}) and (\ref{rhok})) using the kink density of an elementary step $\rho_{k,1}$ (Eq. (\ref{rhok1})) and the kink density of a double step $\rho_{k,2}$ (Eq. (\ref{rhok2})).
Using the equations of the kink densities, we reproduced the surface velocities obtained by the Monte Carlo method (\S \ref{MCmovements}, Table \ref{table_kink}) without fitting parameters.

The time lag of the surface motion for an abrupt reverse of the driving force around equilibrium is one of the methods of detection for the step-droplets that are created by the discontinuity in the surface tension.
When $\Delta \mu >0$ (growth) at low temperature, elementary steps separate successively from the lower side of the step edge on a macro-step (Fig. \ref{growth} (b), (c)).
The elementary steps in the terrace of the upper side of the macro-step catch up and merge with the macro-step (Fig. \ref{growth} (d), (e)). 
Hence, the profile of a macro-step is asymmetric with respect to the upper side and the lower side of the step edge. 
Then, when the driving force is reversed abruptly to $\Delta \mu<0$ (sublimation), the movement of the elementary steps reverses; for the macro-step, however, the profile of the step edges changes first.
In this way, a time lag of the surface motion occurs.
If step droplets do not exist, as in the original RSOS model, the bunched steps dissolve rapidly because the edges of the side surface of the bunched step are rough near equilibrium.   
There is no nucleation barrier for the separation of steps from the bunched steps. 
Therefore, the inhibition of the macro-step motion against the alternative change of $\Delta \mu$ around $\Delta \mu = 0$ becomes the evidence of the discontinuity in the surface tension.

It should be noted that under equilibrium, step merging or step faceting do not always occur for sticky steps.
In order for step merging to occur under equilibrium, discontinuities in the surface tension and in the vicinal surface free energy are essential.
For temperatures higher than $T_{f,1}$, in the case of the p-RSOS model, discontinuities disappear because the entropic repulsion between the steps overwhelms the short-range step-step attraction.
Step faceting then disappears.
The step-droplets, on the other hand, remain for temperatures slightly above $T_{f,1}$, due to the finite character of the step-droplets. 
Therefore, a slowing down of the step velocity occurs when $|\Delta \mu|$ is small.

Recently, an anomaly together with strong anisotropy in the stiffness of vicinal surfaces was observed\cite{parshin11} for the surface around the (0001) facet of a $^4$He crystal particle at low temperature.
This anomaly in the stiffness of the $^4$He crystal particle may be explained by anomalous behavior of the surface stiffness tensor that originated from a discontinuity in the surface tension (\S \ref{stiffness},\S \ref{faceting}).
Based on this present work and our previous work, we consider that short-range step-step attraction with a microscopic origin might exist on the surface of a $^4$He crystal.
A hydrodynamic interaction of steps\cite{uwaha90} may also be conceivable for $^4$He.
Since the hydrodynamic interaction of steps is long range, it causes a discontinuity in the surface tension of the vicinal surface tilted in all directions around the (0001) facet at a temperature lower than the specific temperature $T_{f}$.
In the case of the point-contact-type step-step attraction, however, the attraction causes a discontinuity in the surface tension only around the vicinal surface tilted towards a special direction where the step-droplets are formed.
In the p-RSOS model, the vicinal surface tilted towards the $[100]$ direction from the (001) surface shows typical GMPT behavior\cite{akutsuJPCM11}.
We expect further relevant experimental studies in the future.

In the present Monte Carlo simulation, only non-conserved attachments and detachments of atoms\cite{hohenberg} were taken into consideration in order to show clearly the effect of the point-contact-type step-step attraction.
That is, other effects that occur on a real surface were ignored, such as surface diffusion\cite{bcf}, electromigration\cite{stoyanov}-\cite{uwaha}, the Schwoebel effect\cite{pimpinelli,weeks}, the shockwave effect\cite{chernov}, impurity effects\cite{weeks94}-\cite{frank}, strain effects\cite{teichert}-\cite{ibach}, and the effect of surfactants\cite{vonhoegen98,minoda}.
On real surfaces, these effects exist together with discontinuities in the surface tension.
For example, for a system with short-range attraction and long-range repulsive step-step interaction, such as an elastic interaction, Tersoff {\it et al.}\cite{tersoff95}, Liu {\it et al.}\cite{liu98}, and Shenoy {\it et al.}\cite{shenoy98} showed that a regular array of merged steps of size $n$ appears on the vicinal surface, where $n$ depends on the strength of the step-step attraction and on the strength of the elastic repulsion.

The vicinal surface of Cu (11n) (n= 5,7,9)\cite{neel} may be an example of a surface with a discontinuity in the surface tension together with an Ehrlich-Schw\"{o}bel (ES) barrier.
The STM image (Fig. 2 (a)) of the Cu (11n) vicinal surface in the paper of N\'{e}el {\it et al.} looks similar to the pattern shown in Fig. \ref{pinning} of the present work.
The dark spots in their STM image, which look like holes or trenches, seem to pin the macro-steps.
An explanation for the pinned phenomena was not given in their paper.
We consider that the trenches may be formed by the self-pinning caused by the step faceting in the early stage of step bunching.
The authors said that the step bunching had a kinematical cause because the regular train of steps was observed at about 700 K.
If the characteristic temperature $T_{f,1}$ is lower than 700 K, however, the macro-steps caused by the sticky character of the steps dissolve at $T \sim 700$ K.

In our previous papers\cite{akutsu03,akutsu09-2}, we presented lattice models to describe the vicinal surface with adsorption.
We showed that the discontinuity in the surface tension and the vicinal surface free energy is induced by adsorbates.
Hence, the inhibition of step motion around equilibrium, as presented in this work, is expected to occur on a vicinal surface with adsorbates.
In fact, the shape shown in Fig. \ref{growth} and intermittent growth, as is one shown in Fig. \ref{intermittent}, are similar to those seen in the observations of a Au/Si(001) surface\cite{minoda} and an O/Ni(977) surface\cite{pearl}.
In the case of the Au/Si(001) surface, gold-induced faceting occurs, and the origin of the faceting is thought to be discontinuities in the surface tension\cite{minoda2}. 

In the case of the O/Ni(977) surface, oxygen-induced step merging has been observed, and it has also been observed on the surfaces of several other metals\cite{ozcomert}.
So far, step bunching, or step merging, has been considered to be a dynamical phenomenon that takes place when the surface is far from equilibrium\cite{misba10},\cite{stoyanov}-\cite{frank},\cite{khare}.
In their studies of step bunching resulting from various causes when the system is far from equilibrium, the surface free energy and the step free energy are assumed to have the GMPT universal form (Eq. (\ref{frho})).
We think that the oxygen-induced step merging that occurs near equilibrium may be explained by the discontinuity in the surface tension that is induced by adsorbates.
For a full understanding of step dynamics on the vicinal surface in combination with step-step attraction and other surface effects, further study is required.

\section{Conclusion}
Point-contact-type step-step attraction causes discontinuities in the surface tension $\gamma_{\rm surf}(\vec{n})$ and in the vicinal surface free energy $f(\vec{p})$ around the  (001) and (111) surfaces at low temperature.
Due to these discontinuities, `step faceting' occurs, which is where the side surface of a macro-step becomes smooth.
Step faceting inhibits the continuous motion of macro-steps under a small driving force $\Delta \mu$.
Step faceting also induces intermittent motion of the surface in the long-time behavior, and it induces the pinning of steps even if there are no impurities, adsorbates, or defects on the surface.

For temperatures such that the discontinuity in the surface tension occurs only around the (111) surface, the slowing down of step movements occurs due to `step-droplets' (boson $n$-mers with finite lifetimes), which are locally merged steps.
The step-droplets roughen the vicinal surface, but they diminish the kink density of the vicinal surface (step smoothing).

\acknowledgements

The author would like to thank Prof. M. Kitamura for valuable discussions.
  This work was supported by a Grant-in-Aid for Scientific Research from the Ministry of Education, Science, Sports and Culture (No. 15540323).

\appendix

\section{Equation of surface motion near equilibrium}\label{mobility}

In this appendix, we derive an equation for the surface motion near equilibrium.

Based on the linear response theory near equilibrium\cite{kuboII}, the time derivative of the height change of the vicinal surface is assumed to equal the variational derivative multiplied by a transport coefficient.
Let $z'(x,y)= z(x,y) + \delta \xi \delta (x,y)$, where $\delta (x-x_0,y-y_0)$ is the delta function.
Then, we write a kinetic equation for the height change of the surface as follows\cite{muller-1}:
\beq
v_z = \frac{\delta \xi}{\delta t}= - {\cal V}(\{ \vec{p}\} ) \frac{\delta G}{\delta \xi},\label{vz}
\eeq
where ${\cal V}(\{ \vec{p}\} )$ is a transport coefficient that depends on the surface gradient and $G$ is the total free energy of the system.
The functional $G$ for $z'(x,y)$ is described as
\beq
G= \int \int {\rm d}x{\rm d}y [f(\vec{p}) - \frac{\Delta \mu}{\Omega}z'(x,y)],
\eeq
where $\vec{p}=(p_1,p_2)=(\partial z'/\partial x, \partial z'/\partial y)$,  and $f(\vec{p})$ represents the surface free energy per $x$-$y$ projected area (vicinal surface free energy).
$f(\vec{p})$ depends on the surface tension $\gamma_{\rm surf} (\vec{n})$ as $f(\vec{p})= \gamma_{\rm surf} (\vec{n})\sqrt{1+p_1^2+p_2^2}$, where $\vec{n}$ represents the unit normal vector at the surface.

Considering $\delta G =0$, where the free energy becomes minimum, at equilibrium, we have
\beq
\frac{\partial}{\partial x^i}\frac{\partial f(\vec{p})}{\partial p_i}+ \frac{\Delta \mu}{\Omega}=0, \quad 
\frac{\partial f(\vec{p})}{\partial \vec{p}}|_{\vec{p}=\vec{p_e}} = \vec{\eta},
\label{ecs2}
\eeq
where $\vec{x}=(x^1,x^2)=(x,y)$.
From the solution of Eq. (\ref{ecs2}), we obtain  
\beq
\vec{\eta}= - [\Delta\mu/(2\Omega)] \vec{x}, \quad 
\tilde{f}(\vec{\eta})=f(\vec{p})|_{\vec{p}=\vec{p}_e}-\vec{\eta} \cdot \vec{p}_e.
\eeq
Then the relation Eq. (\ref{ecs2}) is inverted as
\beq
\vec{p}_e=- \partial \tilde{f}(\vec{\eta})/\partial \vec{\eta}.
\eeq

Expanding $G$ around the equilibrium $\vec{p}=\vec{p}_e$, we write
$\delta G$ as
\beqa
\delta G = \int \int {\rm d}x{\rm d}y \left[ \left. \frac{\partial^2 f(\vec{p})}{\partial p_i \partial p_j}\right|_{\vec{p}= \vec{p}_e}\delta p_i \delta p_j - \frac{\Delta \mu}{\Omega}\delta \xi \right] \label{distribution} \\
= - \int \int {\rm d}x{\rm d}y \left[ \left. \frac{\partial^2 f(\vec{p})}{\partial p_i \partial p_j}\right|_{\vec{p}= \vec{p}_e}\frac{\partial^2 z}{\partial x_i \partial x_j} \delta \xi + \frac{\Delta \mu}{\Omega}\delta \xi \right].
\label{deltaG}
\eeqa
From Eq. (\ref{vz}) and Eq. (\ref{deltaG}), therefore, we have Eq. (\ref{vz2}).

\section{\label{fdtheorem}Transport coefficient based on the simple linear response theory}

Let us consider the Langevin equation for $\Delta \mu=0$ as follows\cite{barabashi}:
\beq
\frac{\partial h}{\partial t}= \nu \nabla^2 h + R(t) , \label{langevin}
\eeq 
where we assume $R(t)$ is a Gaussian white noise such as $\langle R(t) R (t') \rangle = 2D_h\delta (t-t')$.
Then, the distribution function $P(h,t)$ of the Fokker-Plank equation is proportional to 
\beq
\exp \left[ - \int {\rm d}^2 x \frac{\nu}{2D_h} |(\vec{p}-\vec{p_e})|^2 \right]. \label{fokker}
\eeq
Comparing Eq. (\ref{fokker}) with Eq. (\ref{distribution}) at $\Delta \mu =0$, where Eq. (\ref{distribution}) should be the Hamiltonian of the Fokker-Plank equation, we see that $f^{(ij)}/\kBT$ in Eq. (\ref{distribution}) corresponds to $\nu/2D_h$ in Eq. (\ref{fokker}).
Hence, we have 
\beq
{\cal V}(\vec{p})=2D_h/\kBT. \label{nu6}
\eeq
Since $R_j \propto (p_j-p_{j,e})$,  we have
\beq
\langle R_iR_j\rangle= const. \times \kBT (f^{(\alpha \beta)-1})_{ij}/L^2,
\eeq
from which follows
\beq
\det \langle R_iR_j\rangle=const.^2 \times \frac{4 \pi^2 g^2 (W(\vec{n})^2)^2}{L^4 (\ln L)^2}. 
\eeq
Therefore, we approximate ${\cal V}(\vec{p})$ as follows:
\beq
{\cal V}(\vec{p})\approx const. \times \frac{4\pi g W^2(\vec{n})}{\kBT L^2\ln L}. \label{vz6}
\eeq

\section{Kink sites for sublimation}

\begin{figure}[htbp]
\begin{center}
\includegraphics[width=7 cm,clip]{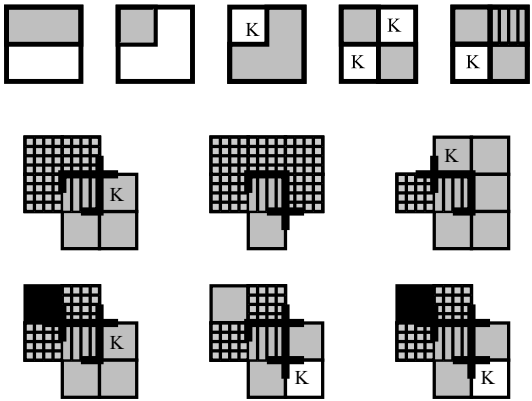}%
\end{center}
\caption{
Conceptual diagrams of the kink configuration for $\Delta \mu<0$ (sublimation).
Top line: kink configurations of an elementary step.
Second and third lines: kink configurations of a double step.
More darkly shaded areas describe the lower terraces: 
white $>$ gray $>$ gray with stripes $>$ gray with hatching $>$ black.
`K' represents a kink site.
}
\label{kink2sublimation}
\end{figure}

In the case of sublimation ($\Delta \mu <0$) at low temperature, `atoms' (elementary cubes in the RSOS model) detach mainly from the kink sites shown in Fig. \ref{kink2sublimation}.
Equations for the kink densities for an elementary step and for a double step are the same as Eq. (\ref{rhok1}) and (\ref{rhok2}).\\


\end{document}